\documentclass[12pt]{article}
\usepackage{amsthm,natbib,url,titlefoot,float,placeins,xr,nolbreaks,alltt,newtxtext,newtxmath,amsfonts,bm,lscape,graphicx,mathtools,multirow}
\allowdisplaybreaks

\newcommand{\blind}{1}

\begin{document}

\def\spacingset#1{\renewcommand{\baselinestretch}%
{#1}\small\normalsize} \spacingset{1}

\newtheorem{definition}{Definition}\newtheorem{hypothesis}{Hypothesis}\newtheorem{lemma}{Lemma}\newtheorem{proposition}{Proposition}\newtheorem{theorem}{Theorem}\newtheorem{corollary}{Corollary}
\newtheorem{assumption}{Assumption}\newtheorem{remark}{Remark}\newtheorem{example}{Example}\newtheorem{condition}{Condition}


\if1\blind
{
  \title{\bf Quantile-Regression Inference With Adaptive Control of Size}
  \author{Juan Carlos Escanciano\thanks{Subsequently published in the {\it Journal of the American Statistical Association} 114(527): 1382--1393.  Juan Carlos Escanciano is Professor and Research Chair in Economics, Universidad Carlos III de Madrid, Calle Madrid 126 28903, Getafe (Madrid), Spain (email:
  jescanci@eco.uc3m.es); and Chuan Goh is Visiting Scholar, Department of Economics and Finance, University of Guelph,
 50 Stone Road East, Guelph, ON, Canada, N1G 2W1 (email: gohc@uoguelph.ca).  This work was partially supported by the Spanish
  Plan Nacional de I+D+I, reference number ECO2014-55858-P.  The authors thank Nicholas Jewell and 
  David Ruppert, an Associate Editor and a referee for comments that greatly improved this paper.  They are also grateful to Victoria 
Zinde-Walsh for her comments and to Daniel Siercks for computing support.}\hspace{.2cm}\\
    Indiana University\\
    and \\
    Chuan Goh\\
    University of Guelph}
  \maketitle
} \fi

\if0\blind
{
  \bigskip
  \bigskip
  \bigskip
  \begin{center}
    {\LARGE\bf Quantile-Regression Inference With Adaptive Control of Size}
\end{center}
  \medskip
} \fi

\begin{abstract}
Regression quantiles have asymptotic variances that depend on the conditional densities of the response variable
given regressors.  This paper develops a new estimate of the asymptotic variance of regression quantiles
that leads any resulting Wald-type test or confidence region to behave as well in
large samples as its infeasible counterpart in which the true conditional
response densities are embedded.  We give explicit guidance on implementing the new variance estimator to
control adaptively the size of any resulting Wald-type test. Monte Carlo evidence indicates the potential of
our approach to deliver powerful tests of heterogeneity of quantile treatment effects in covariates with good size
performance over different quantile levels, data-generating processes and sample sizes. We also include an empirical
example.  Supplementary material is available online.
\end{abstract}

\noindent%
{\it Keywords:}  Regression quantile, asymptotic variance, standard error,
conditional density estimation.
\vfill

\newpage
\spacingset{1.45} 
\section{Introduction}
\label{sec:intro}

\noindent Consider an independent and identically distributed (iid) sample
$(\bm{X}_{1},Y_{1}),\ldots,(\bm{X}_{n},Y_{n})$, where each $Y_{i}$ is scalar-valued, and where, for some
fixed $d$, each $\bm{X}_{i}$ is a $d$-dimensional regressor. We assume that the
conditional distribution of the $i$th response variable $Y_{i}$ given
$\bm{X}_{i}$ satisfies
\begin{equation}
\Pr\left[  \left.  Y_{i}\leq\bm{X}_{i}^{\top}\bm{\beta}(\alpha
)\right\vert \bm{X}_{i}\right]  =\alpha\label{qr}%
\end{equation}
almost surely (a.s.) for some fixed quantile $\alpha\in(0,1)$, where
$\bm{\beta}(\alpha)\in\mathbb{R}^{d}$ is unknown and $\bm{X}_{i}^{\top}$ denotes the transpose of $\bm{X}_{i}$. The
relation (\ref{qr}) specifies a linear $\alpha$-quantile regression model.
Models of conditional quantiles, such as the model given above in
(\ref{qr}), have taken on an important role in the statistical sciences. They generally
offer researchers the possibility of being able to engage in a
systematic analysis of the effects of a set of conditioning variables on all
aspects of the conditional distribution of a response variable. A notable
characteristic of this approach is the ability it gives researchers to model
only the quantiles of interest to a given empirical study without the need to
construct an explicit model for the other regions of the response density. For
example, a researcher may by varying the quantile index $\alpha$ examine the
specific effects of regressors on any point of the conditional distribution of
the response variable. Thus the differential effects of some medical
intervention ($X$) on survival time ($Y$) can be analyzed separately for
low-risk and high-risk individuals by constructing estimates of the
conditional quantile function of $Y$ given $\bm{X}$ for various quantiles.
The monograph of \citet{Koenker05} and the volume edited by
\citet{KoenkerChernozhukovHePeng17} provide comprehensive
reviews of quantile-regression methodology, along with illustrative examples of
its application in various disciplines.

There are several proposals available for quantile regression inference. Some of these proposals,
such as certain methods involving resampling \citep[contains a comprehensive review]{He17}, approaches
based on the asymptotic behavior of regression rank scores \citep{GutenbrunnerJureckova92}, direct methods
\citep{ZhouPortnoy96,FanLiu16} or more recent Bayesian approaches \citep{YangHe12,FengChenHe15,YangWangHe16} differ
from Wald-type methods by avoiding the need to estimate conditional density functions for the purpose of asymptotic
variance estimation of conditional quantile estimators.  Wald-type procedures, however, do generally retain the
attractive feature of computational simplicity, and perhaps for this reason remain popular in empirical practice.

In this paper we develop a new estimator of the asymptotic covariance matrix
of a given regression quantile. The new estimator is explicitly intended to
induce the Wald-type tests or confidence regions in which it is embedded to
behave as well in large samples as their empirically infeasible counterparts
in which the true, as opposed to estimated, conditional densities appear. The
asymptotic variance estimator proposed here induces the empirical size
distortions of Wald-type tests to vanish at the same rate enjoyed by the
corresponding tests incorporating the actual conditional density functions,
i.e., the disparity between the actual and nominal sizes of these tests
vanishes at the \textit{adaptive} rate.

There is of course a long history on estimation of the asymptotic variance of quantile regression parameters
and the corresponding Wald-type tests. Among existing procedures, two implementations that are particularly popular
are those of \citet{Powell91} and \citet{HendricksKoenker92}.  We show that the proposals of \citet{Powell91} and \citet{HendricksKoenker92}
both induce Wald-type tests whose empirical size distortions cannot vanish at the adaptive rates
that become possible when these tests incorporate the asymptotic variance
estimator that we develop below.

The proposed estimator for the conditional density evaluated at the
conditional quantile has applications beyond the formulation of Wald-type tests with
adaptive control of size. This estimator can be used for counterfactual
wage decompositions in a quantile regression setting \citep{MachadoMata05}.
It has been used for developing improved specification tests for linear quantile
regression \citep{EscancianoGoh14}. Semiparametrically efficient
inference in linear quantile regression requires, either explicitly or implicitly, an
estimator of the so-called efficient score, which involves the
conditional density evaluated at the quantile \citep{NeweyPowell90,KomunjerVuong10}.  Finally, estimates of conditional densities
are also needed in semiparametric extensions of the basic linear quantile regression model, e.g., \citet{MaHe16} and
\citet{FengZhu16}. Further applications of our estimator such as these are of independent interest.

Finally, we note that this paper is partly motivated by a recent contribution
of \citet{Portnoy12} to the effect that the first-order asymptotic normal
approximation for regression quantiles is associated with an error bound of
order $O_{p}\left(  n^{-1/2}(\log n)^{3/2}\right)  $. This in turn implies, as
we show below, the benchmark $O_{p}\left(  n^{-1/2}(\log n)^{3/2}\right)$-rate at which size distortions for Wald-type
tests regarding quantile-regression parameters converge when the conditional response
densities are assumed to be known.  An important point to note is that the
error bound of nearly $n^{-1/2}$-order elucidated by \citet{Portnoy12} is smaller
than the error bound of nearly $n^{-1/4}$-order associated with the classic
Bahadur representation for regression quantiles. In particular, the larger
error of nearly $n^{-1/4}$-order is in fact larger in magnitude than the
estimation error associated with any set of reasonable estimates of the
conditional response densities, including those proposed by \citet{Powell91} and
\citet{HendricksKoenker92}. This would apparently suggest that the
rate-adaptive implementation of Wald-type tests proposed in this paper is at
best of second-order importance. The smaller error bound shown by \citet{Portnoy12}
effectively allows one to consider the question of optimally implementing
Wald-type tests in this context as a methodological issue of first-order importance.

The remainder of this paper proceeds as follows. The next section develops the
asymptotic properties of our proposed kernel estimator of the conditional response
density evaluated at the conditional quantile of interest. Section~\ref{bws}
analyzes the size distortions of tests of linear restrictions of quantile
coefficients based on the asymptotic distribution of regression $\alpha$-quantiles. This section also discusses
conditions for our Wald-type tests to exhibit size distortions that decay at the adaptive rate in large samples.
Section~\ref{mc} presents the results of a series of simulation experiments which illustrate the potential of our methods
to deliver accurate and powerful tests, and which are motivated from our
empirical application, which in turn is discussed in Section~\ref{ee}.  An online
supplement includes precise statements of the assumptions underlying our
theoretical results, proofs of those results, additional simulation evidence, details on implementation and further discussion of the 
empirical example.

\section{The New Estimator}

\label{main}

\noindent Consider the $\alpha$-quantile regression model given above in
(\ref{qr}). For each quantile $\alpha\in(0,1)$, the \textit{regression
$\alpha$-quantile} \citep{KoenkerBassett78} is defined as
\[
\hat{\bm{\beta}}_{n}(\alpha)\equiv\arg\min_{\bm{b}\in\mathbb{R}^{d}}%
\sum_{i=1}^{n}\rho_{\alpha}\left(  Y_{i}-\bm{X}_{i}^{\top}\bm{b}%
\right)  ,
\]
where $\rho_{\alpha}(u)=u\left(  \alpha-1\left\{  u\leq0\right\}  \right)$.

For each $i=1,\ldots,n$, let $f_{i}(y)$ and $F_{i}(y)$ denote the conditional
density and cumulative distribution function (cdf), respectively, of $Y_{i}$ given $\bm{X}_{i}$, evaluated at
$y$. If one assumes that for each $i$, $F_{i}(y)$ is absolutely continuous,
and that $f_{i}(y)$ is finite and bounded away from zero at $y=\bm{X}%
_{i}^{\top}\bm{\beta}(\alpha)$, then under Assumption~1 as given in Appendix~A of
the supplementary material, the regression $\alpha$-quantile is asymptotically
normal with
\begin{equation}
\sqrt{n}\left(  \hat{\bm{\beta}}_{n}(\alpha)-\hat{\bm{\beta}}(\alpha
)\right)  \overset{d}{\rightarrow}N\left(  0,\bm{V}(\alpha)\right)  ,
\label{asynorm}%
\end{equation}
where $\bm{V}(\alpha)=\alpha(1-\alpha)\bm{G}_{0}^{-1}(\alpha
)\bm{HG}_{0}^{-1}(\alpha)$ \citep[e.g.,][Theorem~4.1]{Koenker05}, and where
\begin{align}
\bm{G}_{0}(\alpha)  &  =E\left[  f_{i}\left(  \bm{X}_{i}^{\top
}\bm{\beta}(\alpha)\right)  \bm{X}_{i}\bm{X}_{i}^{\top}\right]
;\label{G0}\\
\bm{H}  &  =E\left[  \bm{X}_{i}\bm{X}_{i}^{\top}\right]  .
\label{H}%
\end{align}
Standard Wald-type inferential procedures based on (\ref{asynorm}) naturally
require the estimation of the matrix $\bm{G}_{0}(\alpha)$, which in turn
requires, at least implicitly, the estimation of the conditional density
functions $f_{i}\left(  \bm{X}_{i}^{\top}\bm{\beta}(\alpha)\right)  $
($i=1,\ldots,n$).

We propose an estimator of the conditional response densities $f_{i}\left(
\bm{X}_{i}^{\top}\bm{\beta}(\alpha)\right)$, estimates of which in turn are
used to specify a new estimator of the matrix $\bm{G}_{0}(\alpha)$
appearing in the asymptotic variance of the regression $\alpha$-quantile. The
new estimator of the conditional densities developed here explicitly exploits
the behavior of the fitted conditional $U_{j}$-quantiles $\bm{X}_{i}%
^{\top}\bm{\hat{\beta}}_{n}\left(  U_{j}\right)  $ over a range of
quantiles $U_{1},\ldots,U_{m}$ that are iid realizations from a uniform
distribution on $\mathcal{A}=[a_{1},a_{2}]$. To motivate the new estimator,
note the identity $F_{i}(y)=a_{1}+\int_{a_{1}}^{a_{2}}1\left\{  y-F_{i}%
^{-1}(\alpha)\geq 0\right\}  d\alpha$ for $a_{1}\leq F_{i}(y)\leq a_{2}.$ This
suggests using a smooth approximation of the indicator function, which after
differentiation leads one to the quantity $\left(  a_{2}-a_{1}\right)  \cdot
h^{-1}E\left[  \left.  K\left(  h^{-1}\left(  y-F_{i}^{-1}(U)\right)  \right)
\right\vert \bm{X}_{i}\right]  $, where $K(\cdot)$ is a smoothing kernel
satisfying the conditions of Assumption~2 in the supplementary material and
where $\left.  U\right\vert \bm{X}_{i}\sim Unif[a_{1},a_{2}]$, where
$a_{1}<\alpha<a_{2}$. This quantity should be a good approximation of
$f_{i}(y)$ as $h\rightarrow0,$ where $h>0$ is a scalar smoothing parameter. In
order to avoid numerical integration, we approximate the integral by a finite
sum with $m$ terms. Note that we certainly could take $m=\infty$, but this
would require numerical integration. In what follows, we let both $m$ and the
scalar smoothing parameter $h$ depend on the sample size $n$, with
$m\rightarrow\infty$ and $h\rightarrow 0$ as $n\rightarrow\infty$.

The discussion above leads to the estimator of $f_{i}\left(  \bm{X}%
_{i}^{\top}\bm{\beta}(\alpha)\right)  $ given by
\begin{equation}
\hat{f}_{ni}\left(  \bm{X}_{i}^{\top}\bm{\hat{\beta}}_{n}%
(\alpha)\right)  =\frac{a_{2}-a_{1}}{mh_{m}}\sum_{j=1}^{m}K\left(  \frac
{1}{h_{m}}\bm{X}_{i}^{\top}\left(  \bm{\hat{\beta}}_{n}\left(
U_{j}\right)  -\bm{\hat{\beta}}_{n}(\alpha)\right)  \right)  \label{fhat}%
\end{equation}
for each $i=1,\ldots,n$. The estimators $\hat{f}_{ni}\left(  \bm{X}%
_{i}^{\top}\bm{\hat{\beta}}_{n}(\alpha)\right)  $ given in (\ref{fhat})
are in turn embedded in the following estimator of the matrix $\bm{G}%
_{0}(\alpha)$ as given above in (\ref{G0}):
\begin{equation}
\bm{\hat{G}}_{n}(\alpha)\equiv\frac{1}{n}\sum_{i=1}^{n}\hat{f}_{ni}\left(
\bm{X}_{i}^{\top}\bm{\hat{\beta}}_{n}(\alpha)\right)  \bm{X}%
_{i}\bm{X}_{i}^{\top}. \label{ghat}%
\end{equation}
We are now in a position to state the main result of this section. Define for
$\alpha\in\mathcal{A}$
\begin{equation}
\bm{D}_{nj}(\alpha)\equiv\sqrt{n}\left[  \left(  \bm{\hat{\beta}}%
_{n}\left(  U_{j}\right)  -\bm{\beta}\left(  U_{j}\right)  \right)
-\left(  \bm{\hat{\beta}}_{n}(\alpha)-\bm{\beta}(\alpha)\right)
\right]  , \label{dnj}%
\end{equation}
$\sigma_{K}^{2}\equiv\int_{-1/2}^{1/2}w^{2}K(w)dw$ and $\Vert K\Vert_{2}%
\equiv\sqrt{\int_{-1/2}^{1/2}K^{2}(w)dw}$. In addition, we adopt henceforth
the notation $g^{(k)}(\bm{X})$ to denote the $k$th-order derivative of any
real-valued measurable function $g(\bm{X})$.

\begin{theorem}
\label{ghatconv}

Under Assumptions 1--4 as given in Appendix~A of the supplementary material, and for each
$\alpha\in\mathcal{A}$,
\begin{equation}
\bm{\hat{G}}_{n}(\alpha)=\bm{G}_{0}(\alpha)+\bm{T}_{1nm}%
(\alpha)+\bm{T}_{2nm}(\alpha)+\bm{T}_{3nm}(\alpha)+\bm{R}%
_{nm}(\alpha), \label{ghatexp}%
\end{equation}
where
\begin{align*}
\bm{T}_{1nm}(\alpha)  &  =\sigma_{K}^{2}\cdot\frac{h_{m}^{2}}{2n}%
\sum_{i=1}^{n}f_{i}^{(2)}\left(  \bm{X}_{i}^{\top}\bm{\beta}%
(\alpha)\right)  \bm{X}_{i}\bm{X}_{i}^{\top},\\
\bm{T}_{2nm}(\alpha)  &  =\sqrt{\frac{-\log h_{m}}{mh_{m}}}\cdot\Vert
K\Vert_{2}\cdot\frac{1}{n}\sum_{i=1}^{n}\sqrt{f_{i}\left(  \bm{X}%
_{i}^{\top}\bm{\beta}(\alpha)\right)  }\bm{X}_{i}\bm{X}_{i}^{\top
},\\
\bm{T}_{3nm}(\alpha)  &  =\frac{a_{2}-a_{1}}{nmh_{m}^{2}}\sum_{i=1}%
^{n}\bm{X}_{i}^{\top}\left[  \sum_{j=1}^{m}\frac{1}{\sqrt{n}}%
\bm{D}_{nj}(\alpha)K^{(1)}\left(  \frac{1}{h_{m}}\bm{X}_{i}^{\top
}\left(  \bm{\beta}\left(  U_{j}\right)  -\bm{\beta}(\alpha)\right)
\right)  \right]  \bm{X}_{i}\bm{X}_{i}^{\top}.
\end{align*}
In addition, $\bm{T}_{1nm}(\alpha)=O_{p}\left(  h_{m}^{2}\right)  $,
$\bm{T}_{2nm}(\alpha)=O_{p}\left(  \sqrt{\log h_{m}^{-1}/\left(
mh_{m}\right)  }\right)  $, $\bm{T}_{3nm}(\alpha)=O_{p}\left(
n^{-1/2}\right)  $ and
\begin{align*}
\bm{R}_{nm}(\alpha)  &  =O_{p}\left(  \frac{1}{n}+\frac{1}{n^{3/2}%
h_{m}^{4}}\right)  +o_{p}\left(  h_{m}^{2}+\sqrt{\frac{-\log h_{m}}{mh_{m}}%
}\right) \\
&  =o_{p}\left(  \bm{T}_{1nm}(\alpha)+\bm{T}_{2nm}(\alpha
)+\bm{T}_{3nm}(\alpha)\right)
\end{align*}
as $n\rightarrow\infty$.
\end{theorem}

The terms $\bm{T}_{1nm}(\alpha)$, $\bm{T}_{2nm}(\alpha)$ and
$\bm{T}_{3nm}(\alpha)$ given in the statement of Theorem~\ref{ghatconv}
are the leading second-order terms in an asymptotic expansion in probability,
for a given $\alpha\in\mathcal{A}$, of $\bm{\hat{G}}_{n}(\alpha)$ about
the estimand $\bm{G}_{0}(\alpha)$ . Consider
\begin{equation}
\tilde{f}_{i}\left(  \bm{X}_{i}^{\top}\bm{\beta}(\alpha)\right)
\equiv\frac{a_{2}-a_{1}}{mh_{m}}\sum_{j=1}^{m}K\left(  \frac{1}{h_{m}%
}\bm{X}_{i}^{\top}\left(  \bm{\beta}\left(  U_{j}\right)
-\bm{\beta}(\alpha)\right)  \right)  , \label{ftilde}%
\end{equation}
which defines a natural, but empirically infeasible, kernel estimator of
$f_{i}\left(  \bm{X}_{i}^{\top}\bm{\beta}(\alpha)\right)  $ that
essentially relies on $\bm{\beta}(\alpha)$ and $\bm{\beta}\left(
U_{j}\right)  $, where $j\in\{1,\ldots,m\}$, being known. Then the term
$\bm{T}_{1nm}(\alpha)$ appearing in the statement of
Theorem~\ref{ghatconv} reflects the conditional asymptotic biases given
$\bm{X}_{i}$ of the estimators $\tilde{f}_{i}\left(  \bm{X}_{i}^{\top
}\bm{\beta}(\alpha)\right)  $, defined above in (\ref{ftilde}). The
magnitude of the term $\bm{T}_{2nm}(\alpha)$, on the other hand, is driven
by the conditional variance given $\bm{X}_{i}$ of $\tilde{f}_{i}\left(
\bm{X}_{i}^{\top}\bm{\beta}(\alpha)\right)  $ about
\[
\left(  a_{2}-a_{1}\right)  \cdot E\left[  \left.  h_{m}^{-1}K\left(
h_{m}^{-1}\bm{X}_{i}^{\top}\left(  \bm{\beta}(U)-\bm{\beta}%
(\alpha)\right)  \right)  \right\vert \bm{X}_{i}\right]  .
\]
Lastly, the term $\bm{T}_{3nm}(\alpha)$ corresponds to the error involved
in estimating $\bm{\beta}(\alpha)$ with $\bm{\hat{\beta}}_{n}(\alpha)$.

\section{Wald-Type Tests With Adaptive Control of Size}

\label{bws}

\noindent We consider the empirical sizes of Wald-type tests of hypotheses of
the form
\begin{equation}
H_{0}:\,\bm{R\beta}(\alpha)-\bm{r}=0, \label{h0form}%
\end{equation}
where $\bm{R}$ is a fully specified $(J\times d)$ matrix with rank $J$,
$\bm{r}\in\mathbb{R}^{J}$ is fully specified and $\alpha$ is a fixed
quantile in $\mathcal{A}=\left[  a_{1},a_{2}\right]  $ with $0<a_{1}<a_{2}<1$.
Define the following:
\begin{align}
\bm{\hat{W}}_{n}  &  \equiv\bm{W}_{n}(\bm{\hat{G}}_{n}%
(\alpha)),\label{what}\\
\bm{W}_{0}  &  \equiv\bm{W}(\bm{G}_{0}(\alpha)), \label{w0}%
\end{align}
where for a generic positive definite matrix $\bm{G}$ we define
$\bm{W}_{n}(\bm{G})\equiv(\bm{RG}^{-1}\bm{H}_{n}%
\bm{G}^{-1}\bm{R}^{\top})^{-1}$ and $\bm{W}(\bm{G}%
)\equiv(\bm{RG}^{-1}\bm{HG}^{-1}\bm{R}^{\top})^{-1}$ with
$\bm{H}_{n}=n^{-1}\sum_{i=1}^{n}\bm{X}_{i}\bm{X}_{i}^{\top}$.

Wald-type tests in this context are based on the asymptotic normality of
regression quantiles; as such, attention is naturally directed to the sampling
behavior of asymptotically-$\chi_{J}^{2}$ statistics of the form
$\{n/[\alpha(1-\alpha)]\}(\bm{R\hat{\beta}}_{n}(\alpha)-\bm{r})^{\top
}\bm{W}_{n}(\bm{G}_{n}(\alpha))(\bm{R\hat{\beta}}_{n}%
(\alpha)-\bm{r})$, where $\bm{G}_{n}(\alpha)$ is a consistent
estimator of the matrix $\bm{G}_{0}(\alpha)$. The focus in this section is
on the effect estimation of the matrix $\bm{G}_{0}(\alpha)$ exerts on the
discrepancy between the empirical and nominal sizes of the associated
Wald-type test.

We address the question of whether a Wald-type test of $H_{0}:\,\bm{R\beta
}(\alpha)-\bm{r}=0$ admits the possibility of \textit{adaptive size
control} as $n\rightarrow\infty$. In particular, is it possible to implement
the estimator $\bm{\hat{G}}_{n}(\alpha)$ given above in (\ref{ghat}) in
such a way as to make the discrepancy between the actual size and nominal
level of a Wald-type test of $H_{0}$ vanish at the same rate as the infeasible
test in which the matrix $\bm{G}_{0}(\alpha)$ is actually known? That the
answer to this question is positive can be seen by considering the empirical
size function of a nominal level-$\tau$ Wald test of $H_{0}$. Let $\chi_{J,\tau
}^{2}$ denote the $(1-\tau)$-quantile of a $\chi_{J}^{2}$-distribution, and
let $\bm{Z}(\alpha)\sim N(0,\bm{V}(\alpha))$, where the covariance
matrix $\bm{V}(\alpha)$ is as given above in (\ref{asynorm}). Then one can
combine the asymptotic normality result in (\ref{asynorm}) with
Theorem~\ref{ghatconv} to deduce the following representation of the size
function:
\begin{align}
&  \Pr\left[  \left.  \frac{n}{\alpha(1-\alpha)}\left(  \bm{\hat{\beta}%
}_{n}(\alpha)^{\top}\bm{R}^{\top}-\bm{r}^{\top}\right)  \bm{\hat
{W}}_{n}\left(  \bm{R}\bm{\hat{\beta}}_{n}(\alpha)-\bm{r}\right)
>\chi_{J,\tau}^{2}\right\vert H_{0}\right] \nonumber\\
&  =\Pr\left[  \frac{1}{\alpha(1-\alpha)}\bm{Z}(\alpha)^{\top}%
\bm{R}^{\top}\bm{W}_{0}\bm{RZ}(\alpha)\right. \nonumber\\
&  \left.  >\chi_{J,\tau}^{2}-\frac{1}{\alpha(1-\alpha)}\left(  h_{m}%
^{2}\Lambda_{1n}(\alpha,0)+\sqrt{\frac{-\log h_{m}}{mh_{m}}}\Lambda
_{2nm}(\alpha,0)+\frac{1}{\sqrt{n}}\Lambda_{3nm}(\alpha,0)\right)  \right.
\nonumber\\
&  \left.  -\Theta_{n}(0)-\Xi_{nm}(0)\right]  , \label{powfnexp}%
\end{align}
where $\Lambda_{1nm}(\alpha,0)$, $\Lambda_{2nm}(\alpha,0)$ and $\Lambda
_{3nm}(\alpha,0)$ are $O_{p}(1)$, $\Theta_{n}(0)$ converges to zero at the
same rate as the error committed by the first-order asymptotic approximation
in (\ref{asynorm}), and where $\Xi_{nm}(0)=o_{p}\left(  h_{m}^{2}+\left[  \log
h_{m}^{-1}/\left(  mh_{m}\right)  \right]  ^{1/2}+n^{-1/2}\right)  $. Precise
expressions for $\Xi_{nm}(0)$, $\Lambda_{knm}(\alpha,0)$ ($k=1,2,3$) and
$\Theta_{n}(0)$ are given in (31)--(35) of the supplementary material.

Inspection of (\ref{powfnexp}) indicates that should the matrix $\bm{G}_{0}(\alpha)$ be assumed or in fact be
known by the researcher, then the magnitude of the term $\Theta_{n}(0)$ indicates the rate of convergence of the
size distortion of the infeasible Wald-type test in which $\bm{G}%
_{0}(\alpha)$ is known, i.e., the \textit{adaptive rate of size control} as
$n\rightarrow\infty$.  It follows that the adaptive rate of size
control is determined by the accuracy of the first-order asymptotic normal
approximation for $\sqrt{n}\left(  \bm{\hat{\beta}}_{n}(\alpha
)-\bm{\beta}(\alpha)\right)$.

An important question in this connection is whether the adaptive rate of size
control is so large as to dominate the estimation error associated with any
reasonable estimate of $\bm{G}_{0}(\alpha)$; in this case one might wonder
if there is much point in concerning oneself with a size-optimal
implementation of a given estimator of $\bm{G}_{0}(\alpha)$. This concern
is particularly relevant if the first-order asymptotic normal approximation to
$\sqrt{n}\left(  \bm{\hat{\beta}}_{n}(\alpha)-\bm{\beta}%
(\alpha)\right)  $ is of nearly $n^{-1/4}$-order, as indicated by traditional
analyses of the Bahadur representation for regression quantiles \citep[e.g.,][Theorem~4.7.1]{JureckovaSen96}.
On the other hand, \citet[Theorem~5]{Portnoy12} has recently established that in fact the error associated
with the first-order normal approximation is of nearly $n^{-1/2}$-order, which
is sufficiently small so as not to dominate strictly the estimation error
committed by a typical estimate of $\bm{G}_{0}(\alpha)$ involving local
smoothing. It follows that at least under the conditions imposed by \citet[Theorem~5]{Portnoy12}, the problem of
constructing a size-optimal estimator of
$\bm{G}_{0}(\alpha)$ by choice of a smoothing parameter should be of
primary concern in empirical practice.

We consider an implementation of the estimator $\bm{\hat{G}}_{n}(\alpha)$
given above in (\ref{ghat}) that causes the corresponding Wald-type test of
$H_{0}:\,\bm{R\beta}(\alpha)-\bm{r}=0$ to exhibit adaptive size
control as $n\rightarrow\infty$. The precise conditions on the bandwidth $h_{m}$
and the grid size $m$ are specified in Assumption~3 in Appendix~A of the supplementary
material. These conditions suffice to make the size distortion of the Wald-type test of $H_{0}$ vanish at the adaptive
rate as $n\rightarrow\infty$:

\begin{theorem}
\label{sizeoptbws} Suppose the validity of Assumptions 1--4 as given in Appendix~A of
the supplementary material. Then the corresponding Wald-type test of $H_{0}$ based on $\bm{\hat{G}}_{n}%
(\alpha)$ exhibits adaptive size control as $n\rightarrow\infty$.
\end{theorem}

\noindent The same conditions also cause the Wald-type confidence interval for a
given linear combination of components of $\bm{\beta}(\alpha)$ to have a
level error that vanishes at the rate enjoyed by the corresponding intervals
in which $\bm{G}_{0}(\alpha)$ does not need to be estimated.

Practical recommendations on the implementation of bandwidth parameters and grid sizes that
satisfy the conditions of Theorem~\ref{sizeoptbws} are given in Section~\ref{mc} below and also in
Appendix~D of the supplementary material.  In particular, Wald-type tests embedding
our proposed estimator of $\hat{\bm{G}}_n(\alpha)$ implemented with a fixed (i.e., non-random) bandwidth are exhibited
in Section~\ref{mc} below and in Appendix~E of the supplementary material.
Appendix~D of the supplementary material, on the other hand, derives an
empirically feasible data-driven bandwidth that induces corresponding Wald-type tests to exhibit adaptive size control
as $n\rightarrow\infty$.

Simulation evidence on the finite-sample performance of Wald-type tests implemented
with the data-driven bandwidth are presented in Appendix~E of the supplementary material.

The following corollary is immediate from Theorem~\ref{sizeoptbws} and \citet[Theorem~5]{Portnoy12}:

\begin{corollary}
\label{benchmark}

Suppose the validity of Assumptions~1--4 as given in Appendix~A of the supplementary
material.  Then the following hold as $n\rightarrow\infty$:

\begin{enumerate}
\item The size distortion of the Wald-type test of
$H_{0}:\,\bm{R\beta}(\alpha)-\bm{r}=0$ involving $\bm{\hat{G}}_{n}(\alpha)$ is $O_{p}\left(  n^{-1/2}(\log
n)^{3/2}\right) $; and

\item the level error of the Wald-type confidence interval involving $\bm{\hat{G}}_{n}(\alpha)$ for a
linear combination of the elements of $\bm{\beta}(\alpha)$ is
$O_{p}\left(  n^{-1}(\log n)^{3}\right) $.
\end{enumerate}
\end{corollary}

\noindent Theorem~\ref{sizeoptbws} and Corollary~\ref{benchmark} jointly
establish that in this context the adaptive rate of size control of Wald-type
tests is of nearly $n^{-1/2}$-order, and that a Wald-type test constructed
using the proposed estimator $\hat{\bm{G}}_{n}(\alpha)$ given above in
(\ref{ghat}) can be implemented to exhibit this rate as $n\rightarrow\infty$.

Finally, Appendix~C of the supplementary material shows that the
estimators of $\bm{G}_0(\alpha)$ proposed by \citet{Powell91} and \citet{HendricksKoenker92} cannot
induce Wald-type tests that control size adaptively in large samples.

\section{Numerical Evidence}

\label{mc}

\noindent We present in this section the results of a series of Monte Carlo simulations that are motivated by the empirical question examined in 
Section~\ref{ee}.  These simulations evaluate the performance of Wald-type tests for testing the
heterogeneity of quantile treatment effects \citep[QTEs; see e.g.,][]{Doksum74} in covariates. We
naturally focus attention on the relative performance of Wald-type tests
incorporating our proposed estimator of $\bm{G}_{0}(\alpha )$. We
compare the empirical size and size-corrected power performance of our
tests to those of ten alternative testing procedures available in version~5.35 of the
\url{quantreg} package \citep{Koenker18} for the R statistical computing
environment \citep{RCoreTeam16}. The simulations presented here are all implemented in R; in particular,
we make use of the \url{quantreg} package to generate simulations for each of the competing testing procedures
that we considered.  R code to implement the simulations presented here is included in the supplementary material.

We consider the data-generating process $Y=1+\sum_{j=1}^{4}X_{j}+D+\delta _{a}(U)DX_{1}+F^{-1}(U)$,
where $\{X_{j}\}_{j=1}^{4}$ are iid standard normal and independent of a
treatment indicator $D,$ which follows a Bernoulli distribution with
probability $1/2,$ where $U$ is an independent U[0,1] and where $a\in\mathbb{R}$
denotes the parameter indexing the family of functions $\left\{\delta_a(\cdot):\,a\in\mathbb{R}\right\}$.
In this model the QTE for a given setting of $a$, expressed as a function of a quantile of interest $\alpha$,
is given by $QTE(\alpha )=1+\delta _{a}(\alpha )X_{1}$.

It follows that for a given quantile $\alpha$, a test of the hypothesis
$H_{0}:\delta _{a}(\alpha )=0$ against $H_{1}:\delta_{a}(\alpha )\neq 0$ corresponds
to a test of the homogeneity of the $\alpha$-QTE in $X_{1}$ against the alternative of heterogeneity.  

We set $F$ in the simulations presented here to a standard normal distribution; results in which $F$
denotes a Student-$t$ distribution with three degrees of freedom are given in Appendix~E.3 of
the supplement.  We consider the following specifications of the heterogeneity parameter $\delta _{a}(\alpha )$:

\begin{itemize}
\item Model 1: $\delta _{a}(U)=a$ (pure location).

\item Model 2: $\delta _{a}(U)=a(1+F^{-1}(U))$ (location-scale model).

\item Model 3: $\delta _{a}(U)=(1-5a)G^{-1}(U)-G^{-1}(\alpha )$, with $G\sim
Beta(1,4).$

\item Model 4: $\delta _{a}(U)=2aG^{-1}(U)$, with $G\sim Beta(0.5,0.5).$

\item Model 5: $\delta _{a}(U)=2aG^{-1}(U),$ with $G\sim Beta(2,2).$

\item Model 6: $\delta _{a}(U)=(\sin (2\pi U)-\sin (2\pi \alpha )-2\pi
a)/2\pi .$
\end{itemize}

\begin{figure}
\caption{Heterogeneity parameters for Models 1--6 under  $\alpha$-QTE heterogeneity ($a=1.50$), where 
$\alpha=0.5$}%
\label{deltasa1p5}
\centering
\includegraphics[width=6in,scale=1, keepaspectratio, trim= 0 -7.5cm -0.25cm -1cm]{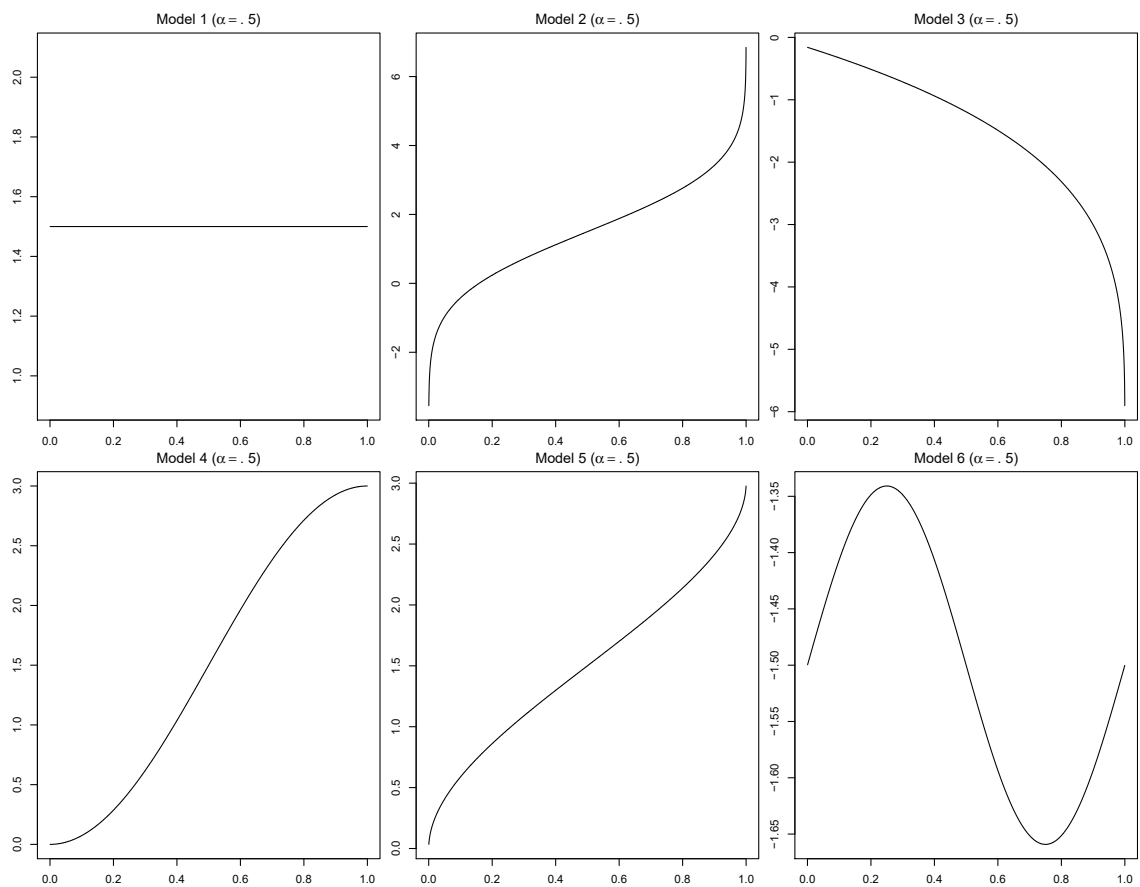}\end{figure}

\noindent Each of these models satisfies the null hypothesis of treatment homogeneity when $a=0$. Under the
null, all models but Models~3 and 6 are pure location models. The alternative
hypothesis corresponds to $a\neq 0.$  Size-corrected power performance is considered against alternatives
corresponding to the settings $a=0.50,$ $1.00$ and $1.50$.  The corresponding heterogeneity parameters for Models 1--6 
under $\alpha=.50$ are plotted in Figure~\ref{deltasa1p5} for the case where $a=1.50$. It is clear that our 
specifications of Models 1--6 imply QTEs with very different functional forms.  

The simulations presented below consider the size and power performance over 1000 Monte Carlo replications
of nominal 5\%-level tests for $\alpha$-quantile regression parameters, where
$\alpha\in\{.25,.50,.75\}$.  Average CPU times over 1000 replications required to implement
each of the tests examined here are also reported.  We considered simulated samples of size $n\in\{100,300\}$.
The techniques used to compute the tests considered are as follows:

\begin{itemize}
\item \url{weg}: Wald-type tests incorporating our proposed
estimator $\bm{\hat{G}}_{n}(\alpha)$, where $\alpha$ is the quantile of
interest. The proposed estimator $\bm{\hat{G}}_{n}(\alpha)$ was
implemented using the Epanechnikov kernel with $m$ quantiles uniformly distributed over the
range $[a_{1},a_{2}]=[.01,,99]$, with
\begin{equation}
m=\left\lfloor \left[  \frac{k}{(\log n)^{\frac{11}{5}}}\right]  ^{\frac
{5}{4}}\right\rfloor \label{mmc}%
\end{equation}
and $k=5$. The bandwidth considered is given by%
\begin{equation}\label{hmmc}
h_{m}=c\left( \frac{\log m}{m}\right) ^{1/5}
\end{equation}
where $c=1.5.$ The choices of $m$ and $h_{m}$ are motivated from the theoretical results presented earlier in Section~\ref{bws}.
The choice of $m$ in (\ref{mmc}) in particular coincides with the lower bound on the rate of divergence
of $m$ as a function of $n$ in our asymptotic results. Appendix~E.1 in the supplement contains
extensive simulation results in which we vary the constants $k$ and $c$. It is shown there that the choice of $k$ is not
as important in terms of finite-sample test performance as the choice of $c$. Our experience with several
data-generating processes, including the ones above, suggest that the choice $c=1.5$ performs very well.
We nevertheless develop in Appendix~D of the supplement a data-driven method for choosing the
bandwidth constant $c$ for a given value of $m$, which is similarly shown in Appendix~E.2 to induce good test performance.

\item \url{riid}: Rank tests assuming a location-shift model with iid errors \citep{Koenker94}.  

\item \url{rnid}: Rank tests assuming a potentially heteroskedastic location-scale-shift model \citep{KoenkerMachado99}.  

\item \url{wiid}: Wald-type tests assuming a location-shift
model with iid errors, with scalar sparsity estimate computed as in \citet{KoenkerBassett78}.  

\item \url{wnid}: Wald-type tests assuming independent but not
identically distributed errors incorporating the difference-quotient estimator
denoted by $\bm{\hat{G}}_{n}^{HK}(\alpha)$ in (38) of the supplement and implemented
using the \citet{HallSheather88} rule-of-thumb bandwidth.  

\item \url{wker}: Wald-type tests assuming independent but not
identically distributed errors incorporating the kernel estimator denoted by
$\bm{\hat{G}}_{n}^{P}(\alpha)$ in (36) of the supplement, where $\bm{\hat
{G}}_{n}^{P}(\alpha)$ was implemented using a uniform kernel supported on
$[-1,1]$ and the bandwidth $\delta_{n}^{P,HS}\equiv\Phi^{-1}\left(
.50+h_{n}^{HS}\right)  -\Phi^{-1}\left(  .50-h_{n}^{HS}\right)  $, where
$h_{n}^{HS}$ is the \citet{HallSheather88} rule-of-thumb bandwidth.  

\item \url{bxy}: Bootstrap tests based on the $(x,y)$-pair method.   

\item \url{bpwy}: Bootstrap tests based on the \citet{ParzenWeiYing94}
method of resampling the sub-gradient condition.  

\item \url{bmcmb}: Bootstrap tests based on the ``MCMB-A'' variant of the
Markov chain marginal bootstrap method of \citet{HeHu02}, described in
\citet{KocherginskyHeMu05}. This variant of the method of \citet{HeHu02}, in
common with the \url{riid} and \url{wiid} methods described
above, assumes an underlying location-shift model with iid errors.  

\item \url{bwxy}: Bootstrap tests based on the generalized bootstrap of
\citet{BoseChatterjee03} with unit exponential weights.  

\item \url{bwild}: Bootstrap tests based on the wild bootstrap method
proposed by \citet{FengHeHu11}.  
\end{itemize}

The Wald-type tests computed using the \url{wiid}, \url{wnid} and
\url{wker} methods were all implemented using the default bandwidth setting in the \url{quantreg} package \citep{Koenker18}, namely the 
\citet{HallSheather88} rule-of-thumb-bandwidth appropriate for inference regarding a population quantile.  In addition, the bootstrap tests were all implemented with the default setting 
of $200$ bootstrap resamples.  

Each of \url{wiid}, \url{wnid}, \url{wker}, \url{bxy}, \url{bpwy}, \url{bmcmb}, \url{bwxy} and \url{bwild} was implemented by direct computation of the 
corresponding test statistic using the corresponding standard error returned by the \url{summary.rq} feature of \url{quantreg}.  On the other hand, the rank-based procedures \url{riid} 
and \url{rnid} both involved direct inversion of the corresponding confidence interval obtained from the \url{summary.rq} feature.

The corresponding simulation results are displayed in Tables~\ref{tabmod1}--\ref{tabmod6}.  These results
include average CPU times in seconds over 1000 replications taken to compute each test statistic.  These average timings
correspond to simulations under the null (i.e., the setting $a=0$) when the quantile of interest is given by $\alpha=0.5$.
Average timings for simulations in which $a\neq 0$ or $\alpha\neq 0.5$ are virtually identical.  

We also examined in unreported work implementations of \url{wiid}, \url{wnid}, \url{wker} and \url{riid} available from the 
\url{anova.rq} feature of \url{quantreg}, but the resulting tests were found to exhibit empirical rejection probabilities that 
were virtually identical to those of the corresponding implementations of these tests using \url{summary.rq}.  We also 
noticed that \url{anova.rq} has a noticeable tendency to run more slowly than \url{summary.rq} for \url{wiid}, \url{wnid} and \url{wker}, and more 
quickly than \url{summary.rq} for \url{riid}.

We see that the empirical size of the proposed method is accurate even with samples of sizes as small as
$n=100$, and is often more accurate than alternative methods, including resampling methods.  We also see that the proposed Wald test has good 
size-corrected power across all six models, three quantiles and two sample sizes for relatively small deviations from the null, i.e. when the constant 
$a$ is small.  It seems clear that an analytical comparison of the asymptotic local relative efficiencies of the different tests 
considered here with that of the asymptotically uniformly most powerful test \citep{ChoiHallSchick96} would be interesting, although 
such an analysis seems beyond the scope of this paper.   We note in passing that the conditional density estimator embedded in our 
method of inference can also be instrumental in estimating the efficient score \citep{NeweyPowell90} and thus in developing asymptotically 
optimal inference for quantile regression.

\FloatBarrier
\begin{table}[H]
\caption{Empirical rejection percentages (size and size-corrected powers) and average execution time, Model 1.  1000 Monte Carlo replications; procedure ``weg'' implemented with fixed bandwidth, $c=1.5$ and $k=5$; other procedures 
implemented using summary.rq.}\label{tabmod1}
\begin{center}
{\scriptsize
\begin{tabular}
[c]{cccccccccccccc}\hline\hline
$n=100$ &  &  & $\alpha=0.25$ &  &  &  & $\alpha=0.5$ &  &  &  & $\alpha=0.75$ & & CPU time ($\alpha=.50$)\\\hline\hline
Method/$a$ & 0 & 0.50 & 1.00 & 1.50 & 0 & 0.50 & 1.00 & 1.50 & 0 & 0.50 & 1.00 & 1.50 & \\\hline
\url{weg} & 5.6 & 16 & 31.6 & 52 & 4.5 & 24.5 & 56 & 81.3 & 5.1 & 21.8 & 43.7 & 68 & 0.0118 \\ 
\url{wiid} & 9.1 & 10 & 22.2 & 39.5 & 7.3 & 15.5 & 45.7 & 75.9 & 8.2 & 12.5 & 31.1 & 56.3 & 0.0025 \\ 
\url{wnid} & 8.1 & 8.3 & 18.7 & 37.9 & 6.8 & 17.5 & 51 & 80 & 7.4 & 12.2 & 33.1 & 59.9 & 0.0021 \\ 
\url{wker} & 1.3 & 13.2 & 31.5 & 53.7 & 0.3 & 17 & 51.2 & 80.8 & 1.9 & 17.7 & 41.8 & 69.5 & 0.0015 \\ 
\url{riid} & 7.9 & 8.6 & 21.4 & 39.4 & 8.6 & 17.7 & 46.5 & 76.9 & 7.5 & 15.3 & 35.5 & 61.5 & 0.0049 \\ 
\url{rnid} & 5.9 & 7.4 & 19 & 37.7 & 6.5 & 17.5 & 46.7 & 76.5 & 5.1 & 15.2 & 34.7 & 61.3 & 0.0156 \\ 
\url{bxy} & 3.1 & 9.6 & 23.6 & 44.7 & 2.9 & 16.7 & 49.8 & 80 & 3.2 & 14.8 & 37 & 65.7 & 0.0212 \\ 
\url{bpwy} & 1.2 & 9.7 & 23.7 & 44.3 & 2.4 & 17.1 & 49.4 & 80.4 & 1.6 & 17.5 & 41.1 & 69.6 & 0.0229 \\ 
\url{bmcmb} & 3.3 & 8.8 & 23.2 & 43.3 & 3.7 & 16 & 48.9 & 79.2 & 3.4 & 16.6 & 39.7 & 66.7 & 0.0137 \\ 
\url{bwxy} & 4.1 & 9.3 & 22.9 & 44.5 & 3 & 16 & 48.4 & 79.9 & 4.4 & 13.7 & 36 & 64.6 & 0.0218 \\ 
\url{bwild} & 6.9 & 10.9 & 24 & 46.2 & 7.2 & 14.1 & 42.7 & 76.1 & 6.2 & 16.2 & 37 & 65.4 & 0.0235 \\ 
$n=300$ &  &  &  &  &  &  &  &  &  &  &  & \\
\url{weg} & 5.4 & 32.1 & 79.8 & 97.7 & 3.2 & 40 & 84.5 & 98.1 & 4.1 & 36.7 & 85.4 & 98 & 0.0453 \\ 
\url{wiid} & 7.9 & 25.4 & 74.2 & 98.1 & 3.7 & 33.6 & 84.3 & 98.5 & 6 & 30.5 & 84.5 & 99.6 & 0.0026 \\ 
\url{wnid} & 8.2 & 26.2 & 76.1 & 98.6 & 3.9 & 34.9 & 86.4 & 98.6 & 5.9 & 32.5 & 84.7 & 99.3 & 0.0035 \\ 
\url{wker} & 3 & 28.4 & 79.5 & 99.3 & 1.3 & 34.5 & 85.9 & 98.7 & 2 & 34.3 & 87 & 99.7 & 0.0017 \\ 
\url{riid} & 7.7 & 27 & 75.8 & 97.6 & 5 & 31.4 & 80.5 & 98.1 & 5.6 & 31.7 & 81.6 & 98.8 & 0.0193 \\ 
\url{rnid} & 6.6 & 26.5 & 74.7 & 97.6 & 4.7 & 31.4 & 80.4 & 98 & 4.7 & 31 & 82.3 & 98.6 & 0.0311 \\ 
\url{bxy} & 4.4 & 29.4 & 79.2 & 98.3 & 2.5 & 34.1 & 84.4 & 98.4 & 3 & 32.7 & 85.5 & 99.4 & 0.0948 \\ 
\url{bpwy} & 3.4 & 28.9 & 78.8 & 98.7 & 2.2 & 34.4 & 84.9 & 98.4 & 2.3 & 34.5 & 85.9 & 99.3 & 0.0991 \\ 
\url{bmcmb} & 5.9 & 26.9 & 77.9 & 98.4 & 3.7 & 33.7 & 82.4 & 98.3 & 3.8 & 32.5 & 84.6 & 99.2 & 0.0369 \\ 
\url{bwxy} & 4.9 & 29.2 & 79.1 & 98.8 & 2.7 & 32 & 82.4 & 98.4 & 3.1 & 31.5 & 83.9 & 99.2 & 0.1002 \\ 
\url{bwild} & 7.1 & 29 & 79.1 & 98.7 & 4.8 & 32.3 & 82 & 98.3 & 4.9 & 31.9 & 85.7 & 99.6 & 0.1018 \\
  \hline
 \end{tabular}
}
\end{center}
\end{table}

\begin{table}[H]
\caption{Empirical rejection percentages (size and size-corrected powers) and average execution time, Model 2.  1000 Monte Carlo replications; procedure ``weg'' implemented with fixed bandwidth, $c=1.5$ and $k=5$; other procedures 
implemented using summary.rq.}\label{tabmod2}
\begin{center}
{\scriptsize
\begin{tabular}
[c]{cccccccccccccc}\hline\hline
$n=100$ &  &  & $\alpha=0.25$ &  &  &  & $\alpha=0.5$ &  &  &  & $\alpha=0.75$ & & CPU time ($\alpha=.50$)\\\hline\hline
Method/$a$ & 0 & 0.50 & 1.00 & 1.50 & 0 & 0.50 & 1.00 & 1.50 & 0 & 0.50 & 1.00 & 1.50 & \\\hline
\url{weg} & 5.7 & 15.4 & 31.3 & 50.5 & 5.9 & 20.6 & 46.5 & 69.6 & 6.4 & 22.2 & 48.8 & 67.5 & 0.0108 \\ 
\url{wiid} & 8.4 & 10.2 & 20.5 & 38.9 & 8.9 & 12.7 & 34.5 & 62.7 & 9 & 15.7 & 39.9 & 63.1 & 0.0022 \\ 
\url{wnid} & 7.4 & 7.3 & 21.8 & 40.2 & 9.1 & 12.6 & 37.2 & 65.3 & 8.6 & 14.5 & 42.1 & 64.4 & 0.002 \\ 
\url{wker} & 1.5 & 8 & 21.9 & 39.9 & 1.1 & 12.5 & 36.5 & 63.2 & 1.7 & 11.3 & 37.7 & 61 & 0.0014 \\ 
\url{riid} & 7.7 & 7.9 & 20.2 & 36.7 & 8.7 & 11.1 & 31.7 & 55.1 & 8.2 & 14.7 & 37.6 & 60.1 & 0.0047 \\ 
\url{rnid} & 5.8 & 7.5 & 20.2 & 36 & 7.2 & 11.4 & 31.1 & 54.1 & 6 & 14 & 35.9 & 57.1 & 0.0142 \\ 
\url{bxy} & 3.4 & 7.9 & 20.3 & 37.6 & 3.4 & 12.6 & 36.2 & 60.3 & 4.1 & 14.6 & 39.3 & 62.1 & 0.021 \\ 
\url{bpwy} & 1.8 & 7.1 & 20.8 & 40.2 & 2.9 & 12.8 & 37.1 & 62.7 & 2.5 & 12.7 & 40 & 62.6 & 0.0225 \\ 
\url{bmcmb} & 3.4 & 8 & 20.5 & 36.7 & 4.1 & 12.7 & 36.2 & 60.1 & 4.6 & 15.3 & 39.2 & 61 & 0.0131 \\ 
\url{bwxy} & 4.5 & 8.3 & 20.6 & 37.9 & 4.2 & 13.2 & 37.1 & 60.2 & 5.2 & 13.5 & 38.7 & 61.5 & 0.0216 \\ 
\url{bwild} & 7.4 & 7.3 & 18.7 & 35.4 & 8.4 & 12.9 & 35.1 & 57 & 7.3 & 14.1 & 38.5 & 59.2 & 0.0229 \\ 
$n=300$ &  &  &  &  &  &  &  &  &  &  &  & \\
\url{weg} & 4.1 & 24 & 64.5 & 88.1 & 3.2 & 41 & 83.9 & 97.7 & 4.9 & 42.5 & 85.4 & 97.3 & 0.0445 \\ 
\url{wiid} & 5.5 & 20.7 & 58.8 & 88.4 & 5 & 32.2 & 81.3 & 98.5 & 8 & 34 & 83.4 & 98.3 & 0.0025 \\ 
\url{wnid} & 5.9 & 19.6 & 60.1 & 88.6 & 4.8 & 35.6 & 84.6 & 98.5 & 8.4 & 36 & 86.5 & 98.9 & 0.0034 \\ 
\url{wker} & 2.3 & 18.4 & 57 & 86 & 1 & 35.9 & 82.2 & 97.9 & 2.3 & 36 & 85.7 & 98.7 & 0.0016 \\ 
\url{riid} & 6 & 17.9 & 55 & 83.9 & 5.4 & 31.8 & 77.7 & 96.5 & 7.5 & 35.6 & 82 & 97.3 & 0.0193 \\ 
\url{rnid} & 4.6 & 17.3 & 53.3 & 83.1 & 5.1 & 30.7 & 76.9 & 96.2 & 6.8 & 33.8 & 80.9 & 96.8 & 0.0311 \\ 
\url{bxy} & 2.6 & 20.7 & 58.8 & 84.2 & 3.7 & 32.7 & 79.9 & 96.9 & 3.7 & 38.4 & 84.6 & 98 & 0.0945 \\ 
\url{bpwy} & 2.4 & 18.1 & 55.3 & 83.7 & 3 & 32.7 & 79.1 & 97 & 3 & 38.5 & 84.7 & 98.4 & 0.0997 \\ 
\url{bmcmb} & 4.3 & 18.3 & 53.1 & 82.7 & 4.4 & 31.2 & 78.4 & 97 & 5 & 37.9 & 84.2 & 97.4 & 0.0369 \\ 
\url{bwxy} & 2.6 & 17.8 & 53.8 & 81.8 & 3.6 & 31.5 & 78.6 & 96.7 & 4 & 36.1 & 82.8 & 97.3 & 0.1003 \\ 
\url{bwild} & 5.1 & 19.1 & 55.8 & 84.1 & 5 & 30.7 & 78.9 & 96.4 & 6.1 & 36.1 & 84.6 & 98.5 & 0.1024 \\ 
\hline
\end{tabular}
}
\end{center}
\end{table}

\begin{table}[H]
\caption{Empirical rejection percentages (size and size-corrected powers) and average execution time, Model 3.  1000 Monte Carlo replications; procedure ``weg'' implemented with fixed bandwidth, $c=1.5$ and $k=5$; other procedures 
implemented using summary.rq.}\label{tabmod3}
\begin{center}
{\scriptsize
\begin{tabular}
[c]{cccccccccccccc}\hline\hline
$n=100$ &  &  & $\alpha=0.25$ &  &  &  & $\alpha=0.5$ &  &  &  & $\alpha=0.75$ & & CPU time ($\alpha=.50$)\\\hline\hline
Method/$a$ & 0 & 0.50 & 1.00 & 1.50 & 0 & 0.50 & 1.00 & 1.50 & 0 & 0.50 & 1.00 & 1.50 & \\\hline
\url{weg} & 5.9 & 12.5 & 24.3 & 43.7 & 4.8 & 21.3 & 43.2 & 63.8 & 5 & 28.3 & 57.6 & 79.4 & 0.0109 \\ 
\url{wiid} & 9.7 & 6.1 & 14.5 & 26.4 & 7.5 & 11 & 28.5 & 53.3 & 7.7 & 16.1 & 44.9 & 71.8 & 0.0023 \\ 
\url{wnid} & 7.9 & 8.4 & 19 & 36.8 & 6.7 & 11 & 31.9 & 56.8 & 7.2 & 18.2 & 47.4 & 72.1 & 0.002 \\ 
\url{wker} & 1.4 & 8.1 & 19.7 & 39.2 & 0.7 & 12.5 & 33.4 & 58 & 1.4 & 18.9 & 52.6 & 78.1 & 0.0014 \\ 
\url{riid} & 7.5 & 6.5 & 15.7 & 32.6 & 7.3 & 9.4 & 26.7 & 47.4 & 8 & 16.9 & 43.9 & 68.2 & 0.0048 \\ 
\url{rnid} & 5.3 & 6.7 & 16.6 & 32.2 & 6.5 & 9.3 & 27.8 & 45.6 & 5.5 & 17.3 & 45.4 & 68.3 & 0.0145 \\ 
\url{bxy} & 2.4 & 8.3 & 19.1 & 37.9 & 2.8 & 12.3 & 32.3 & 55.7 & 3 & 19.3 & 49.2 & 75.2 & 0.021 \\ 
\url{bpwy} & 1.2 & 8.1 & 20.3 & 38.2 & 2.4 & 11.6 & 31.8 & 54.2 & 1.5 & 18.7 & 50.2 & 75.7 & 0.0228 \\ 
\url{bmcmb} & 2.6 & 7.5 & 18.5 & 34.5 & 3.6 & 11.6 & 31.8 & 54.7 & 3.1 & 18.1 & 47.1 & 73 & 0.0133 \\ 
\url{bwxy} & 3.1 & 8.5 & 20.2 & 37.6 & 3.5 & 10.7 & 30.9 & 54.2 & 3.9 & 18.9 & 49.5 & 74.3 & 0.0215 \\ 
\url{bwild} & 6.3 & 7.7 & 18.5 & 35.7 & 7.6 & 10 & 27.7 & 50.2 & 7 & 17.1 & 47.2 & 73.6 & 0.0235 \\ 
$n=300$ &  &  &  &  &  &  &  &  &  &  &  & \\
\url{weg} & 4.9 & 18.3 & 53 & 83.1 & 4.3 & 29.6 & 75.8 & 96.4 & 6.1 & 44 & 88.9 & 98.4 & 0.044 \\ 
\url{wiid} & 6.6 & 12.5 & 46.4 & 81.3 & 6.9 & 24.4 & 74 & 96.3 & 6.9 & 41.1 & 91.1 & 99.5 & 0.0025 \\ 
\url{wnid} & 6.8 & 14.7 & 52.7 & 84.1 & 5.8 & 28.7 & 78.2 & 97.3 & 7.7 & 41.4 & 92 & 99.7 & 0.0035 \\ 
\url{wker} & 3.3 & 15.4 & 52.7 & 84.5 & 1.6 & 28.2 & 76.7 & 96.2 & 3.2 & 40 & 90.4 & 99.7 & 0.0017 \\ 
\url{riid} & 5.8 & 15.6 & 49.7 & 82.2 & 6.4 & 26 & 72.1 & 95 & 7.3 & 38.3 & 87.3 & 98.9 & 0.0193 \\ 
\url{rnid} & 5 & 15 & 48.1 & 80.5 & 6 & 25.4 & 70.4 & 94.4 & 6.4 & 37.9 & 86.5 & 99 & 0.0308 \\ 
\url{bxy} & 3.7 & 16.1 & 50.3 & 83.3 & 3.5 & 27.3 & 74.7 & 95.6 & 3.8 & 41.1 & 89.9 & 99.6 & 0.0946 \\ 
\url{bpwy} & 3.1 & 15.6 & 52 & 83.7 & 3 & 28.2 & 75.4 & 95.9 & 2.8 & 38.5 & 89.8 & 99.2 & 0.0993 \\ 
\url{bmcmb} & 4.7 & 14.8 & 49.7 & 81 & 4.7 & 28.7 & 76.5 & 96 & 5.1 & 40.8 & 90.5 & 99.4 & 0.0367 \\ 
\url{bwxy} & 3.7 & 14.9 & 51 & 82.8 & 3.7 & 28.5 & 75.7 & 96 & 4.2 & 39.9 & 90 & 99.6 & 0.1001 \\ 
\url{bwild} & 6.3 & 13.9 & 48.7 & 81.9 & 5.9 & 25.3 & 73.3 & 95.7 & 6.8 & 37.8 & 88.9 & 99.5 & 0.1021 \\ 
\hline
\end{tabular}
}
\end{center}
\end{table}

\begin{table}[H]
\caption{Empirical rejection percentages (size and size-corrected powers) and average execution time, Model 4.  1000 Monte Carlo replications; procedure ``weg'' implemented with fixed bandwidth, $c=1.5$ and $k=5$; other procedures 
implemented using summary.rq.}\label{tabmod4}
\begin{center}
{\scriptsize
\begin{tabular}
[c]{cccccccccccccc}\hline\hline
$n=100$ &  &  & $\alpha=0.25$ &  &  &  & $\alpha=0.5$ &  &  &  & $\alpha=0.75$ & & CPU time ($\alpha=.50$)\\\hline\hline
Method/$a$ & 0 & 0.50 & 1.00 & 1.50 & 0 & 0.50 & 1.00 & 1.50 & 0 & 0.50 & 1.00 & 1.50 & \\\hline
\url{weg} & 6.5 & 14.2 & 27 & 45.3 & 4.7 & 23.1 & 52.6 & 73.2 & 6.2 & 22.4 & 49.3 & 74.6 & 0.0115 \\ 
\url{wiid} & 9.8 & 6 & 16 & 30.4 & 7.5 & 14.2 & 41.1 & 68.6 & 9.8 & 13 & 36.8 & 66.9 & 0.0025 \\ 
\url{wnid} & 8.5 & 6.3 & 15.6 & 32.2 & 7.8 & 14 & 42.8 & 69 & 8.2 & 15.9 & 43.8 & 71.8 & 0.0021 \\ 
\url{wker} & 1.4 & 11.7 & 24.5 & 43.2 & 1.1 & 12.7 & 43 & 66.9 & 1.7 & 16.2 & 45.1 & 73.2 & 0.0015 \\ 
\url{riid} & 7.7 & 7.4 & 17.8 & 31.6 & 7.4 & 15.3 & 40.3 & 63.1 & 7.9 & 14.4 & 41 & 67.2 & 0.0049 \\ 
\url{rnid} & 5.4 & 8.2 & 18.8 & 34.5 & 6.3 & 13.8 & 39.5 & 62.1 & 5.5 & 15.9 & 41.7 & 68.4 & 0.0154 \\ 
\url{bxy} & 3.2 & 9.1 & 19.6 & 37.8 & 3.6 & 14.6 & 42.5 & 65.3 & 3.1 & 17.6 & 46.3 & 72.6 & 0.021 \\ 
\url{bpwy} & 1.5 & 8.5 & 20.7 & 38 & 2.7 & 13.8 & 40.6 & 64.2 & 1.1 & 17.1 & 47.3 & 74.9 & 0.0234 \\ 
\url{bmcmb} & 4.4 & 6.7 & 17.2 & 33.3 & 4.1 & 14 & 41.2 & 64.2 & 3.2 & 17.3 & 45.9 & 71.4 & 0.0136 \\ 
\url{bwxy} & 4.4 & 8.9 & 20.2 & 37.7 & 3.9 & 15 & 42.9 & 66.2 & 4.3 & 17.7 & 47 & 72.5 & 0.0216 \\ 
\url{bwild} & 7.4 & 9.2 & 20.7 & 37.3 & 6.7 & 13.6 & 40 & 64.3 & 7.8 & 15.3 & 41.5 & 68.7 & 0.0233 \\ 
$n=300$ &  &  &  &  &  &  &  &  &  &  &  & \\
\url{weg} & 4.9 & 24.9 & 59 & 85 & 3.9 & 36.5 & 81.5 & 97.3 & 5 & 45.4 & 87 & 98.5 & 0.0438 \\ 
\url{wiid} & 6.5 & 14.5 & 48.1 & 81.5 & 6.9 & 28.7 & 79.6 & 97.7 & 5.9 & 40.1 & 88.3 & 99.2 & 0.0025 \\ 
\url{wnid} & 7.3 & 17.6 & 53.3 & 84.3 & 7.2 & 28.2 & 79.6 & 97.6 & 5.9 & 42.9 & 90.3 & 99.4 & 0.0034 \\ 
\url{wker} & 3.5 & 23.3 & 59.9 & 87.1 & 2.1 & 28.9 & 78.6 & 97.6 & 2.3 & 41.1 & 88.9 & 98.8 & 0.0016 \\ 
\url{riid} & 7.2 & 17.4 & 49.2 & 81.6 & 8 & 26.7 & 76.5 & 96 & 5.6 & 40.4 & 86.6 & 98.2 & 0.0191 \\ 
\url{rnid} & 6.1 & 17.5 & 50.8 & 81.9 & 6.8 & 25.6 & 76.1 & 95.5 & 4.8 & 41 & 86 & 98.1 & 0.0306 \\ 
\url{bxy} & 4.5 & 18.6 & 52.2 & 82.5 & 3.8 & 27.8 & 77.6 & 96.4 & 3.3 & 40.8 & 87.6 & 98.3 & 0.0937 \\ 
\url{bpwy} & 4 & 18 & 55 & 84.5 & 4.4 & 28.9 & 77.2 & 96.5 & 2.3 & 42.2 & 87.9 & 98.5 & 0.0992 \\ 
\url{bmcmb} & 5.6 & 17 & 50.8 & 81.5 & 5.7 & 28.8 & 78.2 & 96.5 & 4.6 & 41.1 & 87.4 & 98.1 & 0.0367 \\ 
\url{bwxy} & 4.5 & 18.2 & 52.5 & 82.5 & 4.8 & 28.1 & 76.7 & 96.2 & 3.3 & 43.5 & 88.6 & 98.4 & 0.0993 \\ 
\url{bwild} & 6.6 & 17.9 & 53.1 & 82.9 & 6.5 & 25.9 & 75.7 & 96 & 5 & 41.5 & 88.3 & 98.7 & 0.1017 \\  
\hline
\end{tabular}
}
\end{center}
\end{table}

\begin{table}[H]
\caption{Empirical rejection percentages (size and size-corrected powers) and average execution time, Model 5.  1000 Monte Carlo replications; procedure ``weg'' implemented with fixed bandwidth, $c=1.5$ and $k=5$; other procedures 
implemented using summary.rq.}\label{tabmod5}
\begin{center}
{\scriptsize
\begin{tabular}
[c]{cccccccccccccc}\hline\hline
$n=100$ &  &  & $\alpha=0.25$ &  &  &  & $\alpha=0.5$ &  &  &  & $\alpha=0.75$ & & CPU time ($\alpha=.50$)\\\hline\hline
Method/$a$ & 0 & 0.50 & 1.00 & 1.50 & 0 & 0.50 & 1.00 & 1.50 & 0 & 0.50 & 1.00 & 1.50 & \\\hline
\url{weg} & 5.8 & 17.3 & 34.4 & 52.8 & 4.7 & 19.9 & 40 & 62.6 & 6.7 & 17.5 & 39.8 & 64.9 & 0.0109 \\ 
\url{wiid} & 8.5 & 10.9 & 22.7 & 42.1 & 7.2 & 11.5 & 27.7 & 55.6 & 9.6 & 11.3 & 28.9 & 54.6 & 0.0023 \\ 
\url{wnid} & 8.2 & 10.1 & 25.5 & 46.1 & 6.7 & 11 & 32.5 & 59.5 & 8.2 & 10.7 & 31 & 57.4 & 0.002 \\ 
\url{wker} & 1.1 & 13.1 & 30.4 & 54.3 & 0.7 & 12.3 & 34 & 60.5 & 1.5 & 12.8 & 35.9 & 65.2 & 0.0014 \\ 
\url{riid} & 7.3 & 11.1 & 25.7 & 45.9 & 8.1 & 9.4 & 27.2 & 51.8 & 8.4 & 11 & 29.5 & 58.8 & 0.0049 \\ 
\url{rnid} & 5.3 & 11.2 & 26.1 & 45.8 & 7 & 10.7 & 27.3 & 51.9 & 6.2 & 11.7 & 28.9 & 56.8 & 0.0145 \\ 
\url{bxy} & 2.7 & 11.4 & 27.1 & 49.4 & 2.5 & 11.9 & 32.7 & 58.7 & 3.3 & 12.4 & 33.3 & 62.7 & 0.021 \\ 
\url{bpwy} & 1.2 & 12.1 & 28.8 & 50.7 & 2.6 & 12.5 & 33.8 & 60.3 & 2 & 12.1 & 34 & 64 & 0.0231 \\ 
\url{bmcmb} & 2.9 & 10.8 & 27.6 & 47.5 & 3.7 & 11.3 & 31.8 & 59.5 & 3.5 & 11.7 & 32.8 & 59.4 & 0.0134 \\ 
\url{bwxy} & 4.2 & 11.3 & 27.7 & 48.8 & 3.6 & 11.3 & 32.4 & 58.6 & 4.4 & 11.8 & 32.7 & 61.8 & 0.0215 \\ 
\url{bwild} & 6.8 & 12.4 & 26.9 & 47.2 & 7 & 9.7 & 28.6 & 53.9 & 7.4 & 10.5 & 31.8 & 61.2 & 0.0231 \\ 
$n=300$ &  &  &  &  &  &  &  &  &  &  &  & \\
\url{weg} & 5.4 & 26.6 & 71.2 & 94.9 & 4.1 & 34.6 & 78.2 & 96.3 & 4.8 & 40 & 84.1 & 97.6 & 0.0456 \\ 
\url{wiid} & 7.2 & 24 & 66.6 & 94.6 & 6.6 & 25.9 & 73.1 & 96.5 & 6.6 & 33.6 & 83.5 & 99.3 & 0.0027 \\ 
\url{wnid} & 6.9 & 24.5 & 68.4 & 95.6 & 6.3 & 29 & 76.7 & 97.5 & 7 & 37.4 & 86.6 & 99.3 & 0.0036 \\ 
\url{wker} & 2.7 & 26.5 & 72.1 & 96.6 & 1.7 & 30.4 & 77.3 & 97.8 & 2.6 & 38.3 & 87.9 & 99.4 & 0.0017 \\ 
\url{riid} & 6.4 & 20.3 & 63.8 & 91.7 & 5.9 & 25.9 & 72.7 & 95.5 & 6.9 & 33.3 & 82.5 & 98.5 & 0.0193 \\ 
\url{rnid} & 5.4 & 22.2 & 66.5 & 92.9 & 5.5 & 26.7 & 73.3 & 95.5 & 5.7 & 34.1 & 83.8 & 98.6 & 0.0318 \\ 
\url{bxy} & 3.6 & 24.7 & 70.3 & 95.5 & 3.8 & 29.4 & 75.4 & 97.4 & 4 & 34.3 & 84.9 & 99 & 0.0944 \\ 
\url{bpwy} & 3.5 & 23 & 68.2 & 95 & 3.6 & 28.6 & 75.9 & 97.1 & 2.7 & 37.7 & 85.8 & 99.2 & 0.0997 \\ 
\url{bmcmb} & 4.9 & 24 & 68.6 & 95.3 & 5 & 28.1 & 75.5 & 96.9 & 4.6 & 36.2 & 85.4 & 99.3 & 0.0373 \\ 
\url{bwxy} & 4 & 24.4 & 69.7 & 95.7 & 4.1 & 29.2 & 75.7 & 97.1 & 4.1 & 35 & 85.2 & 99.1 & 0.1 \\ 
\url{bwild} & 6.4 & 23.1 & 69.1 & 95.7 & 6 & 28.1 & 74.7 & 97.1 & 5.7 & 35.3 & 85.1 & 99.1 & 0.1026 \\ 
\hline
\end{tabular}
}
\end{center}
\end{table}

\begin{table}[H]
\caption{Empirical rejection percentages (size and size-corrected powers) and average execution time, Model 6.  1000 Monte Carlo replications; procedure ``weg'' implemented with fixed bandwidth, $c=1.5$ and $k=5$; other procedures 
implemented using summary.rq.}\label{tabmod6}
\begin{center}
{\scriptsize
\begin{tabular}
[c]{cccccccccccccc}\hline\hline
$n=100$ &  &  & $\alpha=0.25$ &  &  &  & $\alpha=0.5$ &  &  &  & $\alpha=0.75$ & & CPU time ($\alpha=.50$)\\\hline\hline
Method/$a$ & 0 & 0.50 & 1.00 & 1.50 & 0 & 0.50 & 1.00 & 1.50 & 0 & 0.50 & 1.00 & 1.50 & \\\hline
\url{weg} & 5.2 & 24.5 & 49.4 & 74 & 3.6 & 21.5 & 59.2 & 88.4 & 5.6 & 12.8 & 30.2 & 54.9 & 0.0109 \\ 
\url{wiid} & 9.7 & 12.2 & 31.5 & 57.4 & 7.2 & 13.6 & 44.1 & 80.4 & 10 & 6.7 & 18.7 & 40 & 0.0023 \\ 
\url{wnid} & 7.3 & 16.3 & 39 & 66.6 & 5.9 & 16.3 & 52.4 & 86.7 & 8.1 & 7.8 & 21.8 & 43.7 & 0.002 \\ 
\url{wker} & 1.3 & 20.2 & 47.4 & 75.8 & 0.8 & 16.3 & 53.5 & 89.2 & 2.2 & 8.9 & 26.2 & 52.1 & 0.0014 \\ 
\url{riid} & 8.4 & 15 & 36.8 & 62.3 & 7.5 & 14.7 & 46.5 & 80.5 & 7.8 & 5.7 & 19.8 & 41.2 & 0.0049 \\ 
\url{rnid} & 6.7 & 13.1 & 35 & 60.1 & 5.5 & 15 & 46.8 & 82.8 & 5.6 & 6.2 & 20.8 & 43.7 & 0.0145 \\ 
\url{bxy} & 2.7 & 17.6 & 41.4 & 70.9 & 2.4 & 16.7 & 52.5 & 86.9 & 3.1 & 8.7 & 25.1 & 50.4 & 0.0209 \\ 
\url{bpwy} & 1.5 & 17.7 & 42.7 & 71.8 & 1.9 & 16.9 & 51.1 & 87.2 & 1.7 & 8 & 22.5 & 48.7 & 0.0228 \\ 
\url{bmcmb} & 3.1 & 15.8 & 40.3 & 69.1 & 3.4 & 15.8 & 51.8 & 85.7 & 3.6 & 8.6 & 23.7 & 50.7 & 0.0133 \\ 
\url{bwxy} & 3.9 & 17.7 & 41.7 & 71.5 & 2.9 & 17.5 & 52.7 & 87.3 & 4.2 & 8 & 23.2 & 49.4 & 0.0214 \\ 
\url{bwild} & 6.9 & 16.2 & 40.2 & 70 & 6.7 & 14 & 46 & 83.3 & 7.3 & 7.3 & 21.8 & 46.2 & 0.0231 \\ 
$n=300$ &  &  &  &  &  &  &  &  &  &  &  & \\
\url{weg} & 5 & 46.8 & 86.4 & 98 & 5.2 & 32.2 & 79.9 & 98.4 & 4.2 & 21 & 61.8 & 93.2 & 0.044 \\ 
\url{wiid} & 6.4 & 39.7 & 87.4 & 99.2 & 8.3 & 25.9 & 76.7 & 97.9 & 6.5 & 13.9 & 54.3 & 91.3 & 0.0024 \\ 
\url{wnid} & 6.4 & 42.8 & 89.3 & 99.7 & 8.1 & 26.7 & 78.4 & 98.9 & 6.5 & 16.6 & 60.6 & 94.2 & 0.0034 \\ 
\url{wker} & 3.2 & 43.6 & 91 & 99.7 & 2.4 & 31.5 & 83.4 & 99 & 2.6 & 16 & 60.5 & 94.5 & 0.0016 \\ 
\url{riid} & 6.9 & 39.3 & 86.5 & 99 & 7.5 & 25.2 & 71.8 & 97 & 6.4 & 15.1 & 56.7 & 92 & 0.0194 \\ 
\url{rnid} & 6.1 & 39.7 & 86.1 & 99 & 6.9 & 27.8 & 76.3 & 97.6 & 5.3 & 15.2 & 55.7 & 92 & 0.031 \\ 
\url{bxy} & 3.2 & 43.8 & 89.3 & 99.4 & 4.4 & 30.3 & 80.1 & 98.4 & 3.1 & 16.1 & 59.3 & 93.5 & 0.0945 \\ 
\url{bpwy} & 3.2 & 42.5 & 88.1 & 99.4 & 4.2 & 29.4 & 80.9 & 98.5 & 3.1 & 16.3 & 58.1 & 93.2 & 0.0997 \\ 
\url{bmcmb} & 5.2 & 40.8 & 88.3 & 99.4 & 6.3 & 28.3 & 78.6 & 98.5 & 4.7 & 15.2 & 58.3 & 93.3 & 0.0368 \\ 
\url{bwxy} & 4.6 & 39.8 & 87.7 & 99.5 & 5 & 28.8 & 79.4 & 98.4 & 3.5 & 16.5 & 58.5 & 93.7 & 0.0999 \\ 
\url{bwild} & 5.7 & 39.7 & 87.7 & 99.5 & 6.7 & 29.2 & 80.4 & 98.7 & 6.1 & 14.2 & 56.8 & 93.1 & 0.1025 \\ 
\hline
\end{tabular}
}
\end{center}
\end{table}
\clearpage

\noindent The simulations presented here, along with further simulations reported
in the supplementary material, indicate the potential of Wald-type tests based on our proposed method to deliver
good size accuracy and reasonable power across a range of quantiles and
data-generating processes. These simulations also support the theoretical results presented earlier in
Section~\ref{main} inasmuch as the size accuracy of the test tends to outperform
those of the other Wald-type tests considered over the three different quantiles and six data-generating processes
considered in our simulations.

\section{Empirical Example}

\label{ee}

\noindent We consider the reemployment bonus experiments conducted in Pennsylvania by the United States Department of Labor between July 
1988 and October 1989 \citep{CorsonDeckerDunstanKerachsky92}. This experiment involved the randomized assignment of new claimants for 
unemployment insurance (UI) benefits into one of several treatment groups or a control group. Claimants assigned to the control group were 
handled according to the usual procedures of the unemployment insurance system, while claimants assigned to treatment were awarded cash 
bonuses if they were able to demonstrate full-time reemployment within a specified qualifying period.

The corresponding data were previously analyzed using quantile-regression methods by \citet{KoenkerBilias01} and \citet{KoenkerXiao02}; 
\citet{KoenkerBilias01} also discuss older literature evaluating similar experiments.  We follow \citet{KoenkerXiao02} by focusing solely 
on a single treatment group, which combined with the control group yields a sample of size $n=6384$.  The corresponding dataset is 
publicly available and can be downloaded from \url{http://www.econ.uiuc.edu/~roger/research/inference/Penn46.ascii}.  Claimants for 
unemployment benefits that were assigned to this treatment were offered a bonus equal to six times the usual weekly benefit if they 
secured full-time employment within 12 weeks.  Because approximately 20\% of the subjects were reemployed within one week and another 
20\% were not reemployed within a 26-week follow-up window, \citet{KoenkerXiao02} assume a quantile-regression specification of the form 
$F_{\log T|\bm{X}}^{-1}(\alpha)=\bm{X}^{\top}\bm{\beta}(\alpha)$, where $\alpha\in[.20,.80]$, where $T$ denotes the duration of 
unemployment in weeks and where the regressors contained in $\bm{X}$ include a constant term, an indicator for assignment to treatment 
and the fourteen demographic or socioeconomic control variables listed in \citet[p. 1603]{KoenkerXiao02}.

We depart from the specification of \citet{KoenkerXiao02} by including interactions of the treatment indicator with each of the control 
variables used by these authors.  We also include interactions of the indicator for gender with indicators for race, Hispanic ethnicity and number of 
dependents.    We consider, for a given quantile in the interval $[.20,.80]$, the hypothesis that the treatment interaction 
terms in $\bm{X}$ are jointly insignificant, i.e., that the effect of treatment at a given quantile in $[.20,.80]$ does not vary with any 
of the control variables included in $\bm{X}$.  Appendix~F of the supplementary material presents some additional 
evidence specific to the question of whether the effect of treatment in this context varies by age or by participants' stated expectation of being 
recalled to a previously held job. 

\begin{figure}
\caption{Pennsylvania reemployment bonus experiment: 6384 observations.
$p$-values for pointwise tests of covariate-homogeneity in treatment effect, $\alpha$-quantile regressions, $\alpha\in
\lbrack.20,.80]$. The dotted horizontal line denotes significance at the 10\%
level.}%
\label{fig.ee.F20}
\centering
\includegraphics[width=6in,scale=1, keepaspectratio, trim= 0 -7.5cm -0.25cm -1cm]{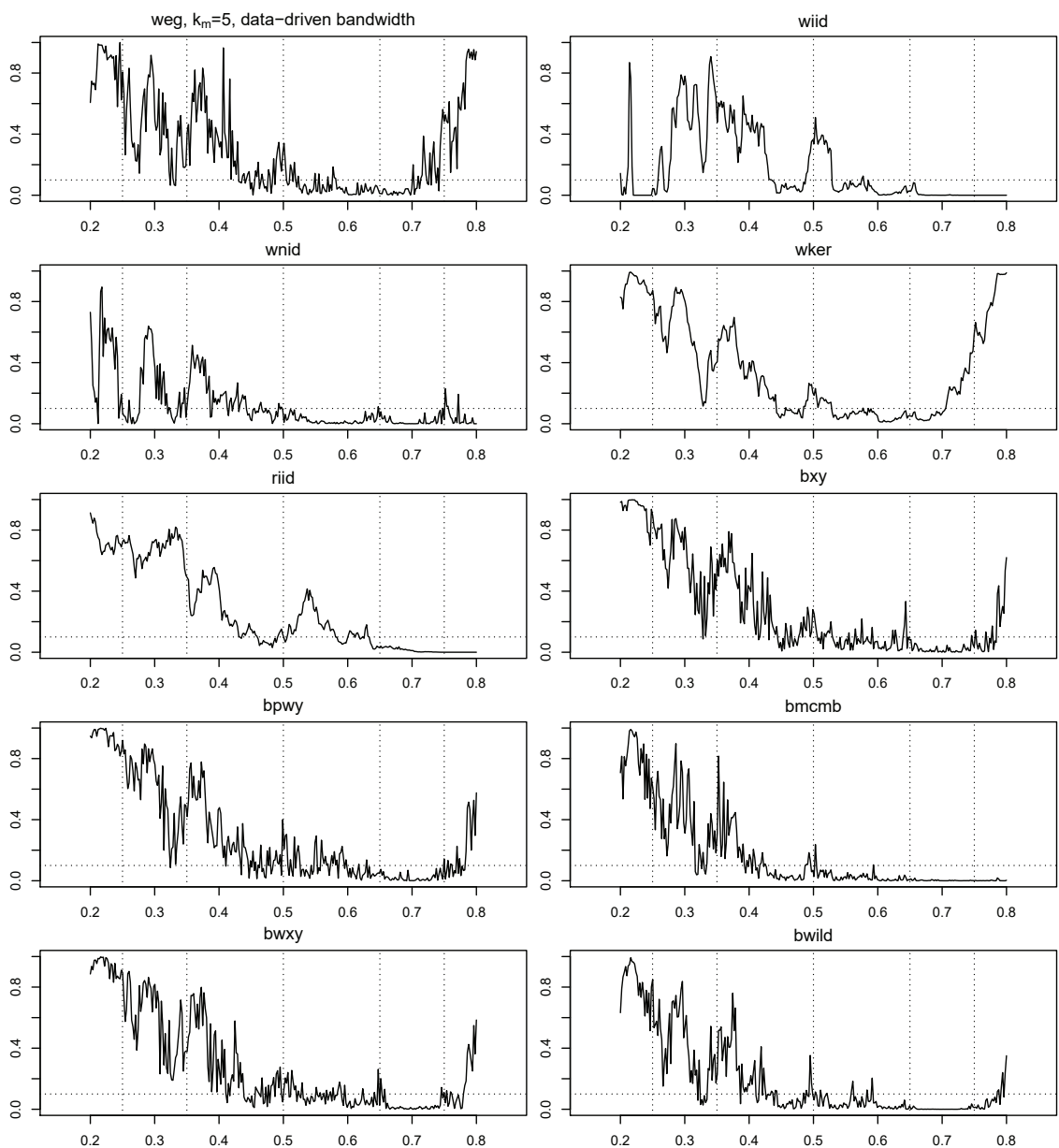}\end{figure}

Figure~\ref{fig.ee.F20} reports $p$-values for the hypothesis of covariate homogeneity in treatment over each quantile in a 
grid of 300 points in $[.20,.80]$.  Our test is implemented using our proposed method with the data-driven bandwidth with $k=5$
discussed in detail in Appendix~D of the supplement.  We also compare the $p$-values from tests implemented using our 
method with the corresponding $p$-values from the alternative testing methods considered in the simulations reported above.  In 
particular, the \url{wiid}, \url{wnid}, \url{wker}, \url{bxy}, \url{bpwy}, \url{bmcmb}, \url{bwxy} and \url{bwild} methods were 
implemented by direct computation of the corresponding Wald-type statistic using the estimated asymptotic covariance matrix generated by 
the \url{summary.rq} feature of version~5.35 of the \url{quantreg} package \citep{Koenker18} for the R statistical computing
environment \citep{RCoreTeam16}.  The \url{riid} method, on the other hand, was implemented by direct invocation of the 
\url{anova.rq} feature of \url{quantreg}.

One can see from Figure~\ref{fig.ee.F20} that our proposed procedure implies significant covariate-heterogeneity in quantile treatment effects at the 
.10-level over nearly all quantiles between .43 and .74.  Unreported results indicate that the joint significance observed at 
these quantiles is driven largely by the significance of two covariates, namely the interaction between 
treatment and an indicator variable for being younger than 35 years of age, and the interaction between treatment and an 
indicator for whether a given participant expected to be recalled to previous employment.  Additional results 
reported in Appendix~F of the supplement reveal significant differences in quantile treatment effects between
participants younger than 35 and those aged 35 and older for nearly all quantiles between .50 and .80.  In particular, the corresponding 
participants aged 35 and older are shown to exit unemployment significantly more slowly than those younger than 35.

Significant differences in quantile treatment effects between participants expecting recall to a previous job and those not expecting 
recall are also shown in Appendix~F to exist for nearly all quantiles between .43 and .74.  This last result is 
potentially important in evaluating the cost-effectiveness of the program given the experiment's exclusion of all claimants for 
unemployment insurance for whom inclusion in the treatment group was deemed not to provide  a sufficient encouragement ``to search for 
work more diligently and to accept suitable employment more rapidly than would be the case otherwise'' 
\citep[p. 10]{CorsonDeckerDunstanKerachsky92}.  The experimenters specifically excluded from the study all claimants who 
indicated a definite expectation of being recalled to a previous employer on a specific date within 60 days of filing their 
applications for UI benefits.  These claimants were deemed to be so secure in their expectation of future full-time employment that any 
bonus paid to them upon resuming full-time employment would be interpreted as a windfall.  Included in the experiment, however, were 
those claimants who indicated some expectation of being recalled to a previous job, although with no definite date of recall.  The 
experimenters deemed claimants in this category to be similar to claimants with no stated expectation of returning to a previous job in 
terms of their assumed response to a promised bonus payment upon resuming full-time employment within the qualifying period.  The
results presented in Appendix~F of the supplement indicate that UI claimants who indicated some expectation of 
being recalled, although not to the extent of having a specific date of recall, in fact differ in their responses to treatment than 
those claimants who indicated no expectation of recall whatsoever.

Figure~\ref{fig.ee.F20} also shows that the other testing methods considered varied in the extent to which the hypothesis 
of covariate-homogeneity in the treatment effect was rejected over quantiles in the interval $[.20,.80]$.  In particular, none of the 
additional inference methods considered was seen to imply the same range of quantiles corresponding to covariate heterogeneity in the 
corresponding quantile treatment effects that was revealed by our method.  For example, \url{wiid} yielded significance at all 
quantiles greater than .53.  We note in addition that some $p$-values for tests implemented using \url{wker} in fact exceed .98 for most quantiles above .78, which suggests that the 
corresponding regression-quantile covariance matrices were not well estimated by \url{wker}.

In view of the rejection, reported by \citet{KoenkerXiao02}, of the null of a linear location-shift model for quantiles on the interval 
$[.25,.75]$, we interpret the \url{wiid} method's conclusion of significance at all quantiles greater than .53 as misleading, and 
likely driven by misspecification of the assumed location-shift model.  As such, inferences resulting from other methods that assume a linear 
location-shift model (i.e., \url{riid} and \url{bmcmb}) are similarly likely to be misleading.  

In summary, we have used our proposed method of inference to show that the effect of treatment on the duration of employment tends to 
vary with individual characteristics of the experimental subjects only over a relatively narrow range of quantiles between .43 and .74.  
These ranges of quantiles corresponding to covariate heterogeneity in the effect 
of treatment is not matched by any of the other testing methods considered.  It 
follows that our proposed method permits an understanding of the effectiveness of a particular unemployment relief policy distinct 
from that produced by other methods of inference.

\bigskip\clearpage
\begin{center}
{\large\bf SUPPLEMENTARY MATERIAL}
\end{center}
\begin{description}
\item[Appendices:] Appendix~A contains precise statements of the assumptions used in
Theorems~\ref{ghatconv} and \ref{sizeoptbws}; Appendix~B contains proofs of
Theorems~\ref{ghatconv} and  \ref{sizeoptbws}; Appendix~C shows that the
estimators of $\bm{G}_0(\alpha)$ proposed by \citet{Powell91} and \citet{HendricksKoenker92} cannot
induce Wald-type tests that control size adaptively in large samples; Appendix~D
describes a data-driven, as opposed to a fixed, bandwidth to implement our proposed estimate of $\bm{G}_0(\alpha)$;
Appendix~E reports further simulation evidence on the finite-sample performance of our
proposed method relative to its competitors, while Appendix~F contains further investigation of the 
empirical example presented in Section~\ref{ee}. (qdf61supp.pdf)
\item[R programs:] We also include R code that enables reproduction of the simulation results in
Section~\ref{mc} and Appendix~E and of the empirical analyses reported in Section~\ref{ee} and Appendix~F.  
(qdf61code.zip)
\end{description}

\bibliographystyle{Chicago}
\bibliography{qdf}

\end{document}